\documentclass[12pt,letterpaper,hyper]{JHEP3}

\usepackage[latin1]{inputenc}
\usepackage{amsmath,epsfig}
\usepackage{amssymb,amsfonts}
\usepackage{graphicx}

\usepackage[active]{srcltx}

\title{Corrections to the statistical entropy of\\ five dimensional black holes }

\preprint{}
\author{
Alejandra Castro$^{~1}$,  and Sameer Murthy$^{~2}$\\
\it $^1${Department of Physics and Michigan Center for Theoretical Physics} \\
\it{University of Michigan}\\
\it{Ann Arbor, MI 48109-1120, U.S.A} \\

\it $^2${Laboratoire de Physique Th\'eorique et Hautes Energies (LPTHE)}\\
\it{Universit\'e Pierre et Marie Curie-Paris 6; CNRS UMR 7589}\\
\it{Tour 24-25, 5$^{\grave{e}me}$ \'etage, Boite 126, 4 Place Jussieu} \\
\it {75252 Paris Cedex 05, France}\\

\textrm{Emails}: \email{aycastro@umich.edu}, \email{smurthy@lpthe.jussieu.fr}\\
}

\abstract{We compute the statistical entropy of the three charge
(D1-D5-p) five dimensional black hole to sub-leading order in a
large charge expansion. We find an agreement with the macroscopic
calculation of the Wald entropy in $R^2$ corrected supergravity
theory. The two calculations have a overlapping regime of validity
which is {\it not} the Cardy regime in the microscopic conformal field theory.
We use this result to clarify
the 4d-5d lift for black holes on Taub-NUT space. In particular, we
compute sub-leading corrections to the formula $S^{4d} = S^{5d}$. In
the microscopic analysis, this correction arises from excitations
bound to the Taub-NUT space. In the macroscopic picture, the
difference is accounted by a mechanism present in a higher
derivative theory wherein the geometry of the Taub-NUT space absorbs
some of the electric charge.}

\renewcommand{\Im}{\mbox{Im}}
\renewcommand{\Re}{\mbox{Re}}

\newcommand{\IR}{\mathbb{R}}
\newcommand{\IZ}{\mathbb{Z}}

\newcommand{\Tr}{\mbox{Tr}}

\newcommand{\CC}{\cal{C}}

\def\s{\sigma}

\def\t{\tau}

\def\CN{{\cal N}}

\def\half{{\frac12}}

\def\CN{{\cal N}}

\def\bea{\begin{eqnarray}}
\def\eea{\end{eqnarray}}
\def\be{\begin{equation}}
\def\ee{\end{equation}}
\def\ba{\begin{align}}
\def\ea{\end{align}}
\def\bse{\begin{subequations}}
\def\ese{\end{subequations}}
\def\1F1{{}_1\!F_1}
\def\2F0{{}_2\!F_0}

\def\CN{{\cal N}}

\def\CC{{\cal C}}

\def\CC{{\cal C}}

\def\Tr{{\rm Tr}}

\font\manual=manfnt
\def\dbend{\lower3.5pt\hbox{\manual\char127}}

\def\p{\partial}

\def\bar{\overline}

\def\CN{{\cal N}}

\def\rt2{\sqrt{2}}
\def\irt2{{1\over\sqrt{2}}}

\def\t{\tilde}

\def\s{\sigma}

\font\cmss=cmss10
\font\cmsss=cmss10 at 7pt
\def\IL{\relax{\rm I\kern-.18em L}}
\def\IH{\relax{\rm I\kern-.18em H}}
\def\rlx{\relax\leavevmode}
\def\ZZ{\rlx\leavevmode\ifmmode\mathchoice{\hbox{\cmss Z\kern-.4em Z}}
 {\hbox{\cmss Z\kern-.4em Z}}{\lower.9pt\hbox{\cmsss Z\kern-.36em Z}}
 {\lower1.2pt\hbox{\cmsss Z\kern-.36em Z}}\else{\cmss Z\kern-.4em
 Z}\fi}

\def\Tr{{\rm Tr}}

\def\rt2{\sqrt{2}}
\def\irt2{{1\over\sqrt{2}}}

\def\t{\tilde}
\def\T{\widetilde}

\def\s{\sigma}

\begin{document}


\section{Introduction and summary}

In the last few years, there has been significant progress
\cite{Behrndt:1998eq,{LopesCardoso:1998wt},{LopesCardoso:1999ur},{LopesCardoso:2000qm},{Ooguri:2004zv},{Sen:2005pu},{Sen:2005iz},{Sahoo:2006pm}}
in computing the entropy of four-dimensional black holes in string
theory beyond the large charge estimate. On the macroscopic side,
the dominant contribution to the entropy is given by
Bekenstein-Hawking formula and the sub-leading corrections are found
by studying higher derivative corrections to the effective action of
string theory. In the presence of such higher derivative effects,
the definition of the thermodynamic entropy is modified and one has
to use the Wald formula
\cite{{Wald:1993nt},{Iyer:1994ys},{Iyer:1995kg}} which generalizes
the Bekenstein-Hawking entropy. For extremal black holes, this can
be summarized elegantly by the entropy function formalism
\cite{Sen:2005wa}.

What made the higher-derivative problem tractable is the
understanding of the off-shell formulation of $\CN=2$ supergravity in
four dimensions and the attractor equations of this theory which make
it simple to find and analyze black hole solutions.  The higher
derivative terms analyzed are packaged as corrections to the
prepotential \cite{Ooguri:2004zv}. In another analysis
\cite{{Sen:2005iz}}, a different combination of the higher derivative
terms -- the Gauss-Bonnet interaction -- was studied and found to
correctly capture the entropy to sub-leading order. It is still not
very clear why only a subset of all possible four derivative
corrections correctly captures the sub-leading entropy.

It is natural to ask whether this analysis can be carried over to
other dimensions. In five dimensions, there has been work
\cite{Hanaki:2006pj} on understanding a certain class of higher
derivative corrections, namely the gravitational Chern-Simons term
and other terms related to it by supersymmetry.  This action was used to find black hole
solutions in
\cite{{Castro:2007hc},{Alishahiha:2007nn},{Castro:2007ci},{Castro:2008ne}}
and corrections to the entropy of five dimensional black holes were
computed. Further references on sub-leading corrections to the five
dimensional solutions in the presence of higher derivative terms
include \cite{{Guica:2005ig},Cvitan:2007hu,Prester:2008iu}.

In this paper, we compute the statistical entropy of a 5d black hole
with a given set of charges to sub-leading order in a large charge
expansion. We find that the sub-leading corrections match those
found by the macroscopic analysis. The black hole we analyze is the
one in which the first accurate microscopic computation of the
leading entropy was done \cite{Strominger:1996sh}, the D1-D5-p black
hole in type IIB string theory on $K3$. The theory has 16 real
supersymmetries and the black hole preserves four of them.

\subsection{Microscopic counting}


For {\it four} dimensional black holes, the exact counting of
microstates beyond the large charge estimate has been achieved  in
$\CN=4$ string theory using the construction of the partition
function for $1/4$ BPS dyons in terms of an auxilliary  mathematical
function called the Igusa cusp form, the unique weight 10 modular
form of $Sp(2,\IZ)$. The degeneracy of states can then be counted by
performing an inverse fourier transform of this function using
contour and saddle point methods. This counting formula was
originally conjectured in \cite{Dijkgraaf:1996it} and then derived in
\cite{{Shih:2005uc},{David:2006yn}} using a D-brane-monopole setup,
and generalized to the counting of all dyons in
\cite{Dabholkar:2008zy}.

The derivation uses the relation of the 4d black holes in question
to a three charge spinning black hole in five dimensions which has
come to be known as the 4d-5d lift  \cite{Gaiotto:2005gf}. The 4d
black holes carry one extra charge which corresponds to a unit KK
monopole at the center of which the 5d black hole is placed. By
making the modulus of the KK circle small or large, the authors of
\cite{Shih:2005uc} then argue that the entropy of the 4d and 5d
black holes are related. More precisely \cite{David:2006yn}, the
microstates of the 4d system can be counted by putting together the
microstates of the 5d system, and the states which are bound to the
KK monopole itself.

It turns out that putting together these two pieces gives a
partition function which is given in terms of the above mentioned
Igusa cusp form $\Phi_{10}$. This function has modular
transformation properties under $Sp(2,\IZ)$ which are much more
powerful than those of $SL(2,\IZ)$ which govern the elliptic genus
of a 2d SCFT. Using these modular transformation properties, one can
systematically deduce the sub-leading corrections to the 4d black
hole entropy \cite{{Jatkar:2005bh}, {David:2006yn}, {Sen:2007qy}}.

For the 5d black hole, the analysis in \cite{Strominger:1996sh} used the related 2d SCFT
$Sym^{Q_1Q_5+1}(K3)$ with $L_0$ eigenvalue equal to the momentum
$n$. The SCFT lives on a circle transverse to the space where the
black hole lives. In this 2d SCFT, one
can apply the Cardy formula to estimate the density of states at
high energies. The Cardy formula is valid for
energies much larger than the central charge, {\it i.e.} $n \gg
Q_1Q_5$. There is a systematic procedure to compute corrections to
the Cardy formula \cite{{Dijkgraaf:2000fq},{Manschot:2007ha}} in the
parameter $\frac{Q_1Q_5}{n}\ll 1$.

On the other hand, in the gravity theory, the configuration looks
like a big 5d black hole when the Schwarzschild radius is much
larger than the string length. In the type II theory on $K3$, this
radius is given by $\frac{R_{Sch}^2}{l_s^2}=\frac{Q_1Q_5}{n}$. One
can now look at finer structures and probe higher derivative
corrections to the black hole entropy; these sigma model corrections
to supergravity will be governed by the small parameter
$\frac{n}{Q_1Q_5}$. This is exactly the opposite regime to the one
above where one can compute corrections to the Cardy formula. One
cannot therefore, naively compare the macroscopic corrections with
the microscopic corrections in the Cardy limit.

One thus needs a new tool to compute the sub-leading expansions of
the statistical entropy in the non-Cardy regime.\footnote{This would not
be necessary if one can map the counting problem to that of finding the
density of states in the Cardy regime of a different CFT. Indeed, as was
observed in \cite{Cvitan:2007hu,Prester:2008iu}, the entropy of the 5d black hole which we consider
can be expressed to subleading order as a Cardy formula of a putative dual SCFT
with $L_0=Q_1$ and $c=6Q_5(n+3)$.
It would be very interesting to understand the microscopic origin of such a SCFT
with these values of charges. We thank the referee for pointing this out. \\
In the remainder of the paper, the phrase "away from the Cardy limit" should be taken to
mean "away from the Cardy limit of any currently understood microscopic SCFT, and in
particular the D1-D5-p SCFT".
} Such a tool can be
found by using the above 4d-5d lift in reverse -- we can rewrite the
5d partition function in terms of the 4d partition function plus
some corrections which physically have to do with the {\it stripping
off} of the modes stuck to the KK monopole. Mathematically, as we
shall see in the following, this is expressed as a precise relation
between the 5d and the 4d partition functions. Having done this, we
can use the powerful mathematical properties of the function
$\Phi_{10}$ to deduce systematically the corrections to the 5d
entropy.

\subsection{Five dimensions v/s four dimensions}

This new tool allows us to understand certain features of 5d
black holes and contrast them against 4d black holes. The first such
feature is spacetime duality. The 4d duality group is bigger than
the 5d one, and in particular it contains the 4d electric-magnetic
duality which is absent in 5d. The manifestation of this duality
which exchanges $n \equiv Q^2 \leftrightarrow P^2 \equiv Q_1 Q_5$
appears through the prepotential  in the 4d gravity theory, to which
worldsheet/membrane instantons (depending on the duality frame) wrapping
the $T^2$ contribute in a crucial way.
These instanton contributions complete the classical linear
prepotential into a transcendental function related to the Jacobi
$\eta$ function which is $S$-duality invariant. The entropy function
which depends on the prepotential is thus also duality invariant.

In five dimensions, one of the circles which these worldsheets/branes wrap
becomes large and the five dimensional supergravity does not see
their effects, and only the contributions $P^2 \gg
Q^2$ are retained. The entropy function as we shall see only contains
the residue of the leading linear piece which is {\it not} duality
invariant, which is consistent since 5d supergravity admits no such
$S$-duality.

Our 5d microscopic counting formula matches the 5d gravity
calculation in this regime of charges $Q_1Q_5 \gg n$; it also agrees
to sub-leading order with the corrections to the Cardy formula using
the Jacobi-Rademacher expansion in the regime $n \gg Q_1Q_5$. The
coefficients of the above two sub-leading corrections are not equal
since they are not related by any duality.

Finally, we use our technique to clear up a slightly confusing point
in the literature having to do with the 4d-5d lift. If we put 5d
supergravity ($+$ corrections) on a background Taub-NUT space of
everywhere low curvature, the theory should still be valid. On such
a space, we can compute the entropy of a black hole sitting at the
center which locally looks 5d. It turns out that there is a subtle
shift in the definition of charge in the 5d theory having to do with
the curvature of the Taub-NUT space, which changes the entropy
expressed in terms of the 4d charges. This small change as we shall
show, agrees precisely with the change computed by the 4d and 5d
microscopic formulas.

The plan of this paper is as follows. In section $\S{\ref{BHin5d}}$,
we present the five dimensional effective theory which arises upon
reduction of type IIB string theory on $K3 \times S^1$. At lowest
order, this is $\CN=4$ supergravity in five dimensions. We then
analyze adding four derivative terms to this action and the
corresponding black hole solutions. In section $\S{\ref{macro}}$, we
discuss the Wald entropy formula in the higher derivative theory. We
then apply it to the rotating BMPV black hole in five dimensions and
present the corresponding corrections to the Bekenstein-Hawking
formula. In section $\S{\ref{micro}}$, we present the microscopic
counting formula and compute the first correction to the large
charge result. In section $\S{\ref{4d5d}}$, we discuss the 4d-5d
lift and the slight difference in 4d and 5d black hole entropies. We
explain this difference through the different mechanisms in the
microscopic and macroscopic understanding. In section
$\S{\ref{conclusion}}$, we summarize our results and suggest future
directions. In the Appendix we briefly sketch the evaluation of the
contour and saddle point integral, some relevant properties of the
Jacobi functions and some details of the Jacobi-Rademacher
expansion.

\section{Black holes in five dimensional supergravity \label{BHin5d}}

In this section we outline the construction of black hole solutions
in the presence of higher derivative corrections, which were
discussed in great detail in \cite{{Castro:2007hc},{Castro:2007ci}}.
The framework is ${\cal N}=2$ supergravity in five dimensions
coupled to $n_V$ vector multiplets. This theory can be embedded in
eleven dimensional supergravity compactified on a Calabi-Yau three
fold, where the lower dimensional theory will depend on the
topological data of CY$_3$. In the remaining sections the discussion
will focus on D1-D5-p system which corresponds to a 1/4 BPS black
hole solutions in ${\cal N}=4$ supergravity. The enhancement of
supersymmetry amounts to choosing CY$_3$=K3$\times$T$^2$ and the
theory has a dual description in type IIB supergravity on
K3$\times$S$^1$.

At the two-derivative level, the effective action is given by
\begin{equation}
S = \frac{1}{4\pi^2} \int d^5 x
\sqrt{g}\left(-R-G_{IJ}\partial_aM^I\partial^aM^J-\frac{1}{2}G_{IJ}F^I_{ab}F^{Jab}
+\frac{1}{24}c_{IJK}A^I_aF^J_{bc}F^K_{de}\epsilon^{abcde}\right)~,
\end{equation}
with $I=1,\ldots n_V+1$ and $a,b=0,\ldots 4$ are tangent space
indices. The scalars $M^I$ can be interpreted as volumes of
two-cycles and $M_I$ the volume of the dual four-cycle, which are
related through the intersection numbers $c_{IJK}$
\begin{equation}
M_{I}=\frac{1}{2}c_{IJK}M^JM^K~.
\end{equation}
In addition the metric of the scalar moduli space is
\begin{equation}
G_{IJ}=\frac{1}{2}\left(M_IM_J-c_{IJK}M^K\right)~.
\end{equation}

The four-derivative corrections of interest are those governed by
the mixed gauge-gravitational Chern-Simons term
\begin{equation}\label{L1}
{\cal L}_1= \frac{c_{2I}}{24\cdot 16}\epsilon_{abcde}
A^{Ia}R^{bcfg}R^{de}_{~~fg}~,
\end{equation}
where $c_{2I}$ is the second Chern class of CY$_3$. The overall
coefficient is determined by the M5-brane anomaly cancelation via
anomaly inflow \cite{Duff:1995wd}. ${\cal L}_1$ by itself is not
supersymmetric, but by using the off-shell formulation of the
supersymmetry algebra one can construct the supersymmetric completion
of (\ref{L1}). As discussed in \cite{Hanaki:2006pj}, these
corrections include all possible terms allowed by the symmetry of the
theory which involve the square of the Riemann tensor. This is true
under the assumption that the hypers decouple from the theory, and
therefore it is consistent to discuss configurations that only
involve Weyl and vector multiplets and  multiplet. 

Taking advantage of the off-shell formalism for the five dimensional
theory, the construction of black holes solutions is greatly
simplified. The simplest way to obtain the corrected solution is by
first imposing BPS conditions, and then utilizing equations of motion
for the gauge field and auxiliary fields.

Backgrounds with unbroken supersymmetry allows for stationary
solutions of the form
\begin{equation}
{ds^2=e^{4U(x)}(dt+\omega)^2-e^{-2U(x)}h_{mn} dx^m dx^n~,}
\end{equation}
where $h_{mn}$ is a 4D hyper-Kahler base space, and the particular case of Taub-NUT is given by
\begin{equation}
h_{mn} dx^m dx^n = \frac{1}{H^0(\rho)} (dx^5 + p^0\cos\theta d\phi)^2 +
{H^0(\rho)}\left(d\rho^2 + \rho^2(d\theta^2+\sin^2\theta
d\phi^2)\right)~,
\end{equation}
with $x^5 \cong x^5 + 4\pi$ and $H^0(\rho)=1+\frac{p^0}{\rho}$. The
rotation is described by $\omega=\omega(x^m)dx^m$, and since we are
interested in corrections to the BMPV black holes we will restrict
the discussion to self dual rotation $d\omega=\star_4d\omega$. For
the Taub-NUT base space this fixes the one-form
\begin{equation}\label{J5D}
\omega =  \frac{J}{8\rho}(dx^5 + p^0\cos\theta d\phi)~.
\end{equation}

The last piece of information from supersymmetry is given by the
variation of the gaugino in the vector multiplet. This results in a
condition between the gauge field to the corresponding scalar field
\begin{equation}
F^I=d(M^Ie^{2U}(dt+\omega))~.
\end{equation}

After exhausting the supersymmetry conditions, the equations of
motion for the explicit action will further determine the full
solution. The variation of the action with respect to the gauge
field, {\it i.e.} Maxwell's equation, results in an exact harmonic
equation
\begin{equation}\label{maxwell}
\nabla^2\left[M_Ie^{-2U}-\frac{c_{2I}}{8}\left((\nabla
U)^2-\frac{1}{12}e^{6U}(d\omega)^2\right)\right]=\frac{c_{2I}}{24\cdot8}\nabla^2\left[2\frac{(\nabla
H^0)^2}{(H^0)^2}-\frac{2}{\rho}\right]~.
\end{equation}
The term to the right of the equality arises from the from curvature
of the base space coupled to the gauge field through
$A^I\wedge\hbox{Tr}(R^2)$,  which behaves as a charge density
governed by the curvature of the base space. Solving (\ref{maxwell})
determines the scalar fields as
\begin{equation}\label{solmax}
M_I(\rho) = e^{2U}\left[ M_I^\infty + \frac{q_I}{4\rho}+
  \frac{c_{2I}}{8}\left((\nabla
U)^2-\frac{1}{12}e^{6U}(d\omega)^2\right)
\right]+e^{-2U}\frac{c_{2I}}{24\cdot4}\left[\frac{(\nabla
H^0)^2}{(H^0)^2}-\frac{1}{\rho}\right] ~,
\end{equation}
with $M_I^\infty$ the value of the moduli at infinity. The constants
$q_I$ are identified with conserved 5d charges by Gauss's law, {\it
i.e.} the integral of the conserved current associated with the
variation of the action with respect to $A^I$. Writing
\eqref{maxwell} as the divergence of the current one can identify the
conserved current. In the absence of dipole charges, this is
equivalent to
\begin{equation}\label{charge}
q_I(\Sigma)=-\frac{1}{4\pi^2}\int_{\partial \Sigma}\star_5 \frac{\partial
{\cal L}}{ \partial F^I}~,
\end{equation}
with $\Sigma$ a spacelike surface and $\partial \Sigma$ is the
asymptotic boundary. As shown in \cite{Castro:2007ci} this integral
is sensitive to the geometry of the base space. For example, the
Taub-NUT geometry interpolates between $\mathbb{R}^4$ at the origin
and $\mathbb{R}^3\times S^1$ at infinity, and dialing the size of
the circle interpolates between a 4d and 5d black hole. Naively one
might expect that (\ref{charge}) is independent of the location of
$\Sigma$, but the delocalized source in (\ref{maxwell}) amounts for
the shift
\begin{equation}\label{shift}
q_I(\Sigma_{\infty})-q_I(\Sigma_0)=-\frac{c_{2I}}{24}~,
\end{equation}
where the 5d electric charge is $q_I(\Sigma_\infty)=q_I$ as defined
in (\ref{solmax}), and $q_I(\Sigma_0)$ corresponds to the 4d
electric charge. Notice that this discrepancy appears after the
inclusion of higher derivatives; for the two-derivative theory the
four and five dimensional charge are equal. We will return to this
shift in section $\S{\ref{4d5d}}$ when discussing the 4d-5d lift.

The only function we haven't specified so far is $U(\rho)$. In the
off-shell formalism, the variation of the scalar auxiliary field
modifies the special geometry constraint and for the solution in
question the equation reads
\begin{equation}\label{MSG}
\frac{1}{6} c_{IJK} M^I M^J M^K=1-\frac{c_{2I}}{24}
\left[e^{2U}M^I\left(\nabla^2U-4(\nabla U)^2+\frac{1}{4}
e^{6U}(d\omega)^2\right)+e^{2U}\nabla^i M^I\nabla_iU\right]~.
\end{equation}
By specifying the internal CY$_3$ manifold and the charge vector
$q_I$, one can iteratively solve non-linear differential equation
for the metric function $U(\rho)$ and fully specify the geometry.

\section{Macroscopic derivation of black hole entropy \label{macro}}

Our discussion focuses on supersymmetric rotating black holes and
the sub-leading corrections to the entropy found in
\cite{Castro:2007ci}. The corrections where found by exploiting the
consequences of the attractor mechanism
\cite{{Ferrara:1995ih},{Ferrara:1996dd},{Ferrara:1996um},{Chamseddine:1996pi}}
and utilizing the entropy function formalism. Here we will briefly
outline the key features of the procedure.

For a semi-classical theory of gravity described by a local action,
the black hole entropy can be obtained as a Noether charge
associated to the diffeomorphism invariance of the theory. This is
the well known Wald's entropy formula
\begin{equation}\label{wald}
S=-\frac{1}{8\pi G_D}\int_{\Sigma}d^{d-2}x\sqrt{h} \frac{\partial {\cal
L}}{\partial R_{\mu\nu\rho\sigma}}\epsilon^{\mu\nu}\epsilon^{\rho\sigma}~.
\end{equation}
For two-derivative gravitational theories this gives
Bekenstein-Hawking area law, {\it i.e.} $S=\frac{A}{4G}$. In
practice (\ref{wald}) is somewhat complicated to manipulate,
specially if we are interested in actions which contain higher
powers of the curvature tensor.

As discussed in \cite{Sen:2005wa} one can reformulate (\ref{wald})
as a Legendre transformation of the action for extremal black holes.
The near horizon geometry of these black holes contain an AdS$_2$
factor which allows one to rewrite (\ref{wald}) as a functional of
the on-shell Lagrangian. Define the Lagrangian density
\begin{equation}
f=\frac{1}{4\pi^2}\int dx^{D-2}\sqrt{g}{\cal L}~.
\end{equation}
The black hole entropy is given by\footnote{ The derivation of
(\ref{S5d}) assumes gauge invariance of the action, which is not
true for Chern-Simons terms in discussion. The resolution is well
understood and we refer the reader to \cite{Sahoo:2006vz} for a
detailed discussion.}
\begin{equation}\label{S5d}
S=2\pi\left(e^I\frac{\partial f}{\partial e^I}+e^0
\frac{\partial f}{\partial e^I}-f\right)
\end{equation}
Here $e^I$ and $e^0$ are potentials associated to electric charge
and rotation, respectively. To evaluate explicitly (\ref{S5d}) we
need the on-shell values of the potentials as a function of the
charges, which is greatly simplified by the attractor mechanism

\subsection{Attractor solution and black hole entropy}

The near horizon geometry of an extremal black hole is governed by
the attractor mechanism. For supersymmetric configurations this
highly constrains the geometry independently of the precise action
one uses. One feature of the attractor is that the values of the
scalar fields at the horizon are fixed by charges carried by the
black hole, independent of initial conditions at infinity.
Additionally, the attractor enhances the solution to be maximally
supersymmetric at the horizon.

The rotating attractor solution in five dimensions is described by a
circle fibered over AdS$_2\times$S$^2$,
\begin{equation}\label{metric}
ds^2=-\ell^2(1-\hat{J}^2)(dx_5+\cos\theta d\phi+e^0\rho
dt)^2+{\ell^2}\left(\rho^2dt^2-\frac{d\rho^2}{\rho^2}\right)-\ell^2d\Omega^2_2~,
\end{equation}
where $\ell$ is the AdS$_2$ radius and $\hat{J}$ is the potential
associated with rotation. For simplicity we set $p^0=1$ in \eqref{metric}. The configuration also holds a 2-form flux
carrying electric charges and the corresponding gauge field is
\begin{equation}\label{vab}
A^I=e^I\rho d\tau-\frac{e^0e^I}{(1+(e^0)^2)}(dx_5+\cos\theta
d\phi+e^0\rho dt)~.
\end{equation}
Here the potentials $e^I$ and $e^0$ are related to the near horizon
fields by
\begin{equation}
e^0=-\frac{\hat{J}}{\sqrt{1-\hat{J}^2}}~,\quad
e^I=\frac{\hat{M}^I}{2\sqrt{1-\hat{J}^2}}~.
\end{equation}

Both (\ref{metric}) and (\ref{vab}) are solely determined by solving
the off-shell BPS conditions, which assures that the background is
an exact solution even after including higher derivative
corrections. The next step is to relate the charges $(q_I,J)$ with
the potentials $(\hat{M}^I, \hat{J})$ and the geometry governed by
the scale $\ell$. The modified special geometry constraint
(\ref{MSG}) relates $\ell$ with the potentials,
\begin{equation}\label{spegeom}
\ell^3=\frac{1}{8}\left(\frac{1}{6}c_{IJK}\hat{M}^I\hat{M}^J\hat{M}^K-\frac{1}{12}c_{2I}\hat{M}^I(1-2\hat{J}^2)\right)~.
\end{equation}

By construction, the five dimensional rotation (\ref{J5D}) is
defined as
\begin{equation}
\hat{J}=\frac{1}{8\ell^3}J~,
\end{equation}
and after using (\ref{spegeom}) we have
\begin{equation}\label{attrJ}
J=\left(\frac{1}{6}c_{IJK}\hat{M}^I\hat{M}^J\hat{M}^K-\frac{1}{12}c_{2I}\hat{M}^I(1-2\hat{J}^2)\right)\hat{J}~.
\end{equation}

Electric charges are defined as a conserved quantity associated to
the variation of action with respect to the corresponding gauge
field. Evaluating (\ref{solmax}) at the horizon, the electric charge
$q_I$ is related to the potentials by
\begin{equation}\label{attrQ}
q_I=\frac{1}{2}c_{IJK}\hat{M}^J\hat{M}^K-\frac{1}{8}c_{2I}\left(1-\frac{4}{3}\hat{J}^2\right)~.
\end{equation}
Both (\ref{attrJ}) and (\ref{attrQ}) are obtained by taking the near
horizon limit of the equations of motion.

Given the attractor geometry (\ref{metric})-(\ref{vab}), one can
evaluate the full action including higher derivative corrections to
compute the entropy function (\ref{S5d}). After some effort, the
semi-classical black hole entropy reads
\begin{equation}\label{finalS}
S^{5d}=2\pi\sqrt{1-\hat{J}^2}\left(\frac{1}{6}c_{IJK}\hat{M}^I\hat{M}^J\hat{M}^K+\frac{1}{6}c_{2I}\hat{M}^I\hat{J}^2\right)
\end{equation}
The next step would be to write the potentials $\hat{M}^I$ and
$\hat{J}$ as a function of charges and rotation by solving
(\ref{attrJ})-(\ref{attrQ}), which would allow us to write
$S^{5d}=S^{5d}(q_I,J)$. For generic intersection numbers $c_{IJK}$
this can only be done perturbatively, but as we will discuss below
the equations are invertible for specific CY$_3$ manifolds.

\subsection{Black holes on K3$\times$T$^2$}

We are interested in corrections to the entropy of 1/4 BPS black
holes in ${\cal N}=4$ with internal manifold CY$_3$=K3$\times$T$^2$.
In the eleven dimensional language, the electric charges $q^I$
correspond to M2-branes wrapping two-cycles. Equivalently, we can
consider type IIB string theory on $K3 \times T2$ and D1-D5-P
charges. The D1-D5 system is extended in the $K3$ and the effective
string extends along one of circles $S^1$ of the $T^2$. The momentum
$P$ is excited along the circle $S^1$.

For CY$_3$=K3$\times$T$^2$, $\hat{M}^1$ denotes the modulus on the
torus and $\hat{M}^i$ the moduli on K3, with $i=2,\ldots 23$. The
non-trivial intersection numbers and second Chern class are
\begin{equation}
c_{1ij}=c_{ij}~, \quad c_{2,1}=c_2(K3)=24~.
\end{equation}
For this specific manifold, equations (\ref{attrJ}) and
(\ref{attrQ}) are invertible allowing to write $(\hat{M}^I,\hat{J})$
in terms of $(q_I,J)$
\begin{eqnarray}\label{potential}
\hat{M}^1&=\sqrt{\frac{1}{2}q_iq_jc^{ij}+\frac{4J^2}{(q_1+\frac{c_2}{24})^2}}{(q_1+\frac{c_2}{8})}~,\\
\hat{M}^i&=c^{ij}q_j\sqrt\frac{(q_1+\frac{c_2}{8})}{\frac{1}{2}q_iq_jc^{ij}+\frac{4J^2}{(q_1+\frac{c_2}{24})^2}}~,\\
\label{potential1} \hat{J}&=\frac{J}{q_1+\frac{c_2}{24}}
\sqrt\frac{(q_1+\frac{c_2}{8})}{\frac{1}{2}q_iq_jc^{ij}+\frac{4J^2}{(q_1+\frac{c_2}{24})^2}}~,
\end{eqnarray}
where we define $c^{ij}$ as the inverse of $c_{ij}$. Inserting
(\ref{potential})-(\ref{potential1}) in (\ref{finalS}), the entropy
as function of charges becomes
\begin{equation}
S=2\pi\sqrt{\frac{1}{2}q_iq_jc^{ij}\left(q_1+\frac{c_2}{8}\right)
-\frac{\left(q_1-\frac{c_2}{24}\right)\left(q_1+\frac{c_2}{8}\right)}{\left(q_1+\frac{c_2}{24}\right)^2}J^2}~.
\end{equation}
Expanding to first order in $c_2$ gives
\begin{equation}\label{correcS}
S=2\pi\sqrt{Q^3-{J}^2}\left(1+\frac{3}{2}\frac{Q_1Q_5}{
Q^3-{J}^2}+\ldots\right)~,
\end{equation}
where we identified the IIB charges as
\begin{equation}\label{Q3J2def}
Q_1Q_5=\frac{1}{2}c^{ij}q_iq_j~,\quad n=q_1~,\quad
Q^3-J^2=Q_1Q_5n-J^2~.
\end{equation}
If all the charges $q_i,J$ scale equally, the expression to
sub-leading order is:
\begin{equation}\label{correcSagain}
S=2\pi\sqrt{Q_1Q_5n}\left(1+\frac{3}{2n} - \frac{J^2}{2Q_1Q_5n}+\ldots\right)~,
\end{equation}
where the sub-leading dependence on angular momentum is due to
leading supergravity result. The higher derivatives terms give rise
to corrections proportional to $J$ as displayed in \eqref{correcS},
but are not important in this regime.

Summarizing, we have an expression for the sub-leading corrections
to the entropy (\ref{correcSagain}) for rotating five dimensional
black holes. These corrections come from the supersymmetric
completion of $c_{2I}A^I\wedge\hbox{Tr}(R^2)$. The macroscopic entropy (\ref{correcSagain}) is what we would like
to compare with the microscopic counting formula.

\section{The microscopic degeneracy formula \label{micro}}

The 5d counting problem of the D1-D5 system on $K3$ is captured by a $(4, 4)$ two-dimensional superconformal field theory along  the worldvolume $\IR \times S^1$ with target space $\mathrm{Sym}^{Q_1Q_5 +1}(K3)$ \cite{Vafa:1995bm}. We denote this sigma model SCFT by
\begin{equation}\label{scft5d}
  X^{5d} = \s(\mathrm{Sym}^{Q_1Q_5 +1}(K3) )~.
\end{equation}
Two of the charges $Q_1, Q_5$ that the black hole carries appear in
the definition of the sigma model. The third charge momentum $n$ and
the angular momentum $l$ appear as the eigenvalues of the
hamiltonian $L_0$ and R-charge $J_0/2$ of the sigma model. The charges
are related to the number of D-branes in the following fashion
\be\label{chnorel} Q_5=N_5~, \quad Q_1 = N_1 - N_5~, \ee
because there is an effective negative unit one-brane charge
generated by the five-brane wrapped on the $K3$. The relevant object
which captures the BPS states is the elliptic genus
\be\label{defEll} \chi (X^{5d}; q,y)  \equiv  \Tr^{X^{5d}}_{RR}
(-1)^{ J_0 - \T  J_0} q^{L_0} \T q^{\bar L_0} y^{ J_0}  \equiv
\sum_{n,l} c^{5d}(Q_1Q_5, n,l) q^n y^l ~. \ee
To estimate the growth of the coefficients of this SCFT, we can use
Cardy's formula and spectral flow in the SCFT
\be\label{Cardyestimate}
\Omega  \sim  \exp\left(\sqrt{\frac{c}{6}L_0-J^2} \right) + \ldots ~
\ee
Plugging in
\begin{equation}\label{defL0J}
c=6Q_1Q_5~,\quad L_0=n~,\quad J^2=\frac{l^2}{4}~.
\end{equation}
we get
\be\label{estimate}
\Omega(Q_1,Q_5,n,l)  \sim \exp(2 \pi \sqrt{Q_1Q_5 n - l^2/4}) + \ldots ~,
\ee

The approximation (\ref{estimate}) is valid at high values of $L_0$,
{\it i.e.} $n \gg Q_1Q_5$. One can actually systematically compute
corrections to this result using an exact formula which determines
the fourier coefficients of the elliptic genus of a symmetric
product SCFT in terms of the fourier coefficients of the original
SCFT (in this case $K3$) \cite{Dijkgraaf:1996xw}. The formula relies
on the modular transformation properties of the elliptic genus under
$SL(2,\IZ)$ and uses the Jacobi-Rademacher expansion
\cite{{Dijkgraaf:2000fq},{Manschot:2007ha}}. By its nature, it is
expressed as a series of corrections to the Cardy formula and can be
used as above when $L_0 \gg c$, {\it i.e.}  $ n \gg Q_1 Q_5$.

On the other hand, the black hole entropy function is valid for
large values of charges when {\it all} the charges scale equally,
i.e. $Q_1 Q_5 \gg n \gg 1$. In order to meaningfully compare the two
expressions, we would need to re-sum the Farey tail expansion in
$Q_1 Q_5/n$ and reexpress it as an expansion in $n/Q_1Q_5$, which
{\it a priori} seems to be a difficult problem.

However, we can make progress using the relation of the elliptic
genus of the symmetric product to the Siegel modular form
$\Phi_{10}$. This is known as the Igusa cusp form and is the unique
weight $10$ modular form of $Sp(2, \IZ)$.  Using the more powerful
Siegel modular transformation properties and a saddle point
approximation, we can compute the expansion of the above elliptic
genus for any regime of charges, in particular  $n/Q_1 Q_5 \ll 1$.
Physically, this is related to the 4d-5d lift which we shall discuss
in a following section. In this section, we shall simply use this
relation to our calculational advantage.

The generating function of the elliptic genus of the symmetric
product is given by \cite{Dijkgraaf:1996xw}
\be\label{dmvvform} Z(\rho, \s, v) \equiv  \sum_{k=0}^{\infty} p^k
\chi (Sym^k(X); q,y) = \prod_{n>0,m \ge 0, l} \frac{1}{(1-p^n q^m
y^l)^{c(nm,l)}}~, \ee
where we have set
\be\label{defpqy} q = e^{2 \pi i \rho}~, \quad p = e^{2 \pi i \s}~,
\quad y = e^{2 \pi i v}~, \ee
and the coefficients $c(n,l)$ are defined through
\be\label{defcnl} \chi(X;q,y) = \sum_{n,l} c(n,l) q^n y^l ~. \ee
For $X=K3$, this generating function is related to the Igusa cusp
form $\Phi_{10}$ as \cite{Dijkgraaf:1996it},
\be\label{ZIgusarel} Z(\rho,\s,v) =
\frac{f^{KK}(\rho,\s,v)}{\Phi_{10}(\rho,\s,v)}, \ee where
\bea\label{deffKK} f^{KK}(\rho,\s,v) & = & p \, q \, y \,
(1-y^{-1})^2 \prod_{m=1}^\infty (1-q^m)^{20}   (1-q^m y)^2 (1-q^m
y^{-1})^2  \; \cr & = & p \, \eta^{18}(\rho) \,
\vartheta_1^2(v,\rho) ~. \eea

We are interested in the microscopic degeneracy of the system with
charges $(Q_1,Q_5,n,l)$, which is given by the coefficient $c(n,l)$
of the sigma model (\ref{scft5d}). This can be expressed as an
inverse Fourier transform of the generating function $Z(\T \rho,\T
\s,\T v)$
\begin{equation}\label{invfive}
   \Omega^{5d}(Q_1,Q_5,n,l) = \oint_{\CC}
   d \T \rho d \T \s d \T v \, e^{-2 i\pi \left( \T \rho n +  \T \s (Q_1 Q_5 +1) +  l \T v \right) }\,
   {Z(\T \rho, \T \s, \T v)}~ .
\end{equation}
The contour $\CC$ in the above integral is presented in appendix
$\S{\ref{integral}}$. In the 4d theory, the choice of contour was
important for the analysis of BPS decays and the associated walls of
marginal stability. These decays happened precisely when the contour
crossed a pole related to the decay. These effects did not affect
the power series expansion for the entropy, but were exponentially
small corrections in the degeneracy formula.

In five dimensions, it is expected from a supergravity analysis
 that there are no such decays corresponding to real codimension one
walls \cite{Bena:2005ay}. Note in this
context that the purely $v$ dependent factors in the function
$f^{KK}$ which have a zero at $v = 0$. These poles therefore do not
exist in the 5d partition function. It would be interesting to
analyze in more detail all the poles of the partition function in
the 5d theory. However, for the purpose of computing power law
corrections to the entropy our analysis is sufficient.

\subsection{Saddle point approximation}\label{sec:saddle}

We can solve the integral (\ref{invfive}) in two steps as in
\cite{{Jatkar:2005bh}, {David:2006yn}, {Sen:2007qy}}. First, we
notice that the dominant pole of the expression $1/\Phi_{10}(\T
\rho,\T \s, \T v)$ is {\it not} factored out by the function
$f^{KK}(\T \rho, \T \s, \T v)$. We can therefore do a contour
integral around this pole and the residue is an integral over two
remaining coordinates. This can be approximated by the saddle point
method to give an asymptotic expansion. We follow the method of
\cite{Jatkar:2005bh, David:2006yn} of which we present some relevant details in
appendix $\S{\ref{integral}}$. The actual evaluation only relies on the
fact that the charges $n, Q_1 Q_5, l$ are large and not on the relative
magnitude of the two charges.\footnote{This fact was also used for computing the
four dimensional black hole entropy in a region where $Q^2,P^2$ are large,
and one was much larger than the other \cite{David:2006yn}.} 

We are interested in the answer to first order beyond the large charge limit, and to this order it is given by 
\be\label{sstat} S^{5d}_{stat}  = S_0 + S_1 ~,\ee
which is to be evaluated at its extremum. The classical ($S_0$) and first correction to the large charge limit ($S_1$) are
\bea\label{S1S0} S_0 & = & - 2 \pi i {\T \rho} n - 2 \pi i \T \s
(Q_1 Q_5 +1) + 2 \pi i (\frac{1}{2} - \t v) l ~,\\\label{S1} S_1 & =
& 12 \ln \T \s - \ln \eta^{24}(\rho) - \ln \eta^{24} (\s) + \ln
f^{KK}(\T \rho, \T \s, \T v)~, \eea
with
\be \T \rho = \frac{\rho \s}{\rho + \s}, \quad \T \s = -
\frac{1}{\rho + \s}, \quad \T v = \frac{1}{2} - \sqrt{\frac{1}{4} +
\T \rho \T \s}~. \ee

Since we are interested in the answer to only the first order beyond
the large charge limit, we can extremize only the classical part
$S_0$ and evaluate the full expression (\ref{sstat}) at those
values. By extremizing the classical functional $S_0$ we obtain
\bea\label{rsvvalues} 
\T \rho & = & \frac{i}{2} \frac{Q_1 Q_5
+1}{\sqrt{Q^3 - J^2}}~, \cr \T \s & = & \frac{i}{2}
\frac{n}{\sqrt{Q^3 - J^2}}~, \eea
where\footnote{The shift of one in $Q_1Q_5$ is not important to
sub-leading order in the black hole regime $Q_1Q_5 >>n$, note the
difference with (\ref{Q3J2def}). This shift will be important in the
Cardy regime  which we discuss below.}
\be\label{defQ3} Q^3 - J^2 \equiv (Q_1 Q_5 +1) n - l^2/4~. \ee
Plugging (\ref{rsvvalues}) in (\ref{S1S0})-(\ref{S1}) gives
\be\label{entropy} S_0(Q_1,Q_5,n)  =  2 \pi \sqrt{Q^3 - J^2} \ee
and
\bea\label{entropy1} S_1(Q_1,Q_5,n) & =&  -\pi \frac{n}{\sqrt{Q^3 -
J^2}}- 24 \ln \eta\left(\frac{l+ i 2\sqrt{Q^3 - J^2}}{2n}\right)   -
24 \ln \eta\left(\frac{-l+ i 2\sqrt{Q^3 - J^2}}{2n}\right) \cr
&  &+18 \ln \eta\left(\frac{i Q_1Q_5}{2\sqrt{Q^3 - J^2}}\right) + 2 \ln
\vartheta_1\left( \frac{1}{2} - \frac{il}{4\sqrt{Q^3 - J^2}},
\frac{iQ_1Q_5}{2\sqrt{Q^3 - J^2}} \right) + \ldots~. \eea

\subsection{Supergravity limit}

In the limit where all the charges  $(n,Q_1,Q_5,l)$ are large and
scale uniformly, we can use the expansion of the functions
$\eta(\tau)$, $\vartheta_1(v,\tau)$  (appendix $\S{\ref{jacobi}}$)
and after dropping higher terms we get
\bea\label{entropy1lim2} S_1(Q_1,Q_5,n)  & = &  4 \pi
\frac{\sqrt{Q^3 - J^2}}{n} -  \pi \frac{Q_1Q_5}{\sqrt{Q^3 - J^2}} +
\ldots \cr & = &  3 \pi \sqrt{\frac{Q_1Q_5}{n}}  + \ldots~ \eea
Combining (\ref{entropy}) and (\ref{entropy1lim2}), the full entropy
formula reads
\be\label{entropylim2} S^{5d}(Q_1,Q_5,n)
 =  2 \pi \sqrt{Q_1 Q_5 n}\left(1 + \frac{3}{2n} - \frac{l^2}{8Q_1Q_5n}  \right) +
 \ldots~
\ee
We see that this agrees with the macroscopic result (\ref{correcSagain})
in the same regime of large charges.

\subsection{Cardy limit}

In the opposite Cardy limit, when $n\gg Q_1Q_5$, and $Q^3 - J^2
\gg1$ we can also expand the result (\ref{entropy1}) to sub-leading
order. In order to do that, we first need to use the modular
transformation properties of the various functions (appendix
$\S{\ref{jacobi}}$)
\bea\label{enmodtrans}
S_1(Q_1,Q_5,n) & = & -  \pi \frac{n}{\sqrt{Q^3 - J^2}}- 24 \ln
\eta\left(\frac{l+ i 2\sqrt{Q^3 - J^2}}{2Q_1Q_5}\right)   - 24 \ln
\eta\left(\frac{-l+ i 2\sqrt{Q^3 - J^2}}{2Q_1Q_5}\right)\cr & & + 18
\ln \eta\left(i\frac{2\sqrt{Q^3 - J^2}}{Q_1Q_5}\right) + 2 \ln
\vartheta_1\left( -i \frac{2\sqrt{Q^3 - J^2}}{Q_1Q_5}
\left[\frac{1}{2} - \frac{i l}{4\sqrt{Q^3 - J^2}}\right],
i\frac{2\sqrt{Q^3 - J^2}}{Q_1Q_5} \right) \cr & &  + 2  \pi
\frac{2n}{\sqrt{Q^3 - J^2}} \left[\frac{1}{2} - \frac{l}{i
4\sqrt{Q^3 - J^2}}\right]^2 + \ldots \eea
Dropping terms of higher order in $Q_1Q_5/n$ in (\ref{enmodtrans})
we get
\bea\label{entropy1lim3} S_1(Q_1,Q_5,n)  & = &  4 \pi
\frac{n}{\sqrt{Q^3 - J^2}} - 3 \pi \frac{\sqrt{Q^3 - J^2}}{Q_1Q_5}
\cr & & \qquad +  \pi \frac{n}{\sqrt{Q^3 - J^2}} -  \pi
\frac{n}{\sqrt{Q^3 - J^2}}  -  \pi \frac{n}{\sqrt{Q^3 - J^2}}  +
\ldots\cr & = & 0 + {\cal O}\left(\frac{1}{\sqrt{Q_1Q_5n-l^2/4}}\right) + \ldots,
\eea
which finally allows us to write the entropy as
\be\label{entropylim3} S^{5d}(Q_1,Q_5,n) =  2 \pi \sqrt{(Q_1 Q_5
+1)n - l^2/4}\left(1 + {\cal O}\left(\frac{1}{Q_1Q_5n -
l^2/4}\right) \right) + \ldots
\ee

Note that unlike in the other limit, all the terms suppressed by
$1/Q_1Q_5$ have dropped away, and the first sub-leading term is
suppressed by $1/(Q^3-J^2)$. This is exactly in agreement with the
more familiar Jacobi-Rademacher expansion to the same order which we
have sketched in appendix $\S{\ref{Radem}}$.

\section{Clarifying the 4d-5d lift \label{4d5d}}

The 4d-5d lift
\cite{{Gaiotto:2005gf},{Elvang:2005sa},{Gaiotto:2005xt},{Bena:2005ni},{Behrndt:2005he}}
is a relation between a black hole in five dimensions carrying three
gauge charges plus angular momentum, and a black hole in four
dimensions carrying the above charges and in addition, a
unit\footnote{This has been extended recently to the case when there
are multiple KK monopoles \cite{Dabholkar:2008zy}.} Taub-NUT charge.
The angular momentum in the five dimensions becomes a mometum along
the Taub-NUT circle at infinity in four dimensions. On application
to a rotating BMPV black hole preserving $1/4$ supersymmetry, the 5d
black hole can be related to a four dimensional $1/4$ dyonic black
hole. This relation can be used to derive an exact counting formula
for $1/4$ BPS dyons in $\CN=4$ string theory \cite{ {Shih:2005uc},
{David:2006yn}}.

As a consequence of the attractor mechanism, the entropy of extremal
black holes is independent of asymptotic value of moduli. By tuning
one of these moduli, one can make the curvature of the Taub-NUT
space large or small. Therefore it seems reasonable to relate the
entropy of 4d dyonic black holes with 5d black holes and the leading
order prescription \cite{Shih:2005uc} was
\be\label{orig4d5d} S^{4d}(Q_1Q_5+1,n,l)= S^{5d}(Q_1Q_5,n,l)~, \ee
This equation however, will receive corrections\footnote{These corrections
are not related to the shift in the charges in (\ref{orig4d5d}).} at sub-leading order
\be\label{new4d5d} S^{4d} (Q_1,Q_5,n,l)= S^{5d} (Q_1,Q_5,n,l)
\left(1 + \frac{c_1}{Q^2} + \ldots \right)~. \ee
The computation of these corrections boils down to computing the
sub-leading corrections to the 5d black hole entropy and comparing
with the known sub-leading corrections to the 4d black hole entropy.
The results of the previous sections fill in this gap, and we can
now explain the origin of the small difference in the 4d and the 5d
black hole entropy both from the microscopic and macroscopic
viewpoints.

In the regime of charges that all the charges are large and scaled
equally, the 5d entropy is (\ref{correcS}),(\ref{entropylim2})
\be\label{5dentropy} S^{5d}(Q_1,Q_5,n,l) = 2 \pi \sqrt{Q_1 Q_5
n}\left(1 + \frac{3}{2n} - \frac{l^2}{8Q_1Q_5n}  \right) + \ldots
\ee
In the same limit, the corresponding 4d black hole with one
additional Taub-NUT charge is [see the review  \cite{Sen:2007qy} and
references therein]
\be\label{4dentropy} S^{4d}(Q_1,Q_5,n,l) = 2 \pi \sqrt{Q_1 Q_5
n}\left(1 + \frac{2}{n} - \frac{l^2}{8Q_1Q_5n} \right) + \ldots \ee

The discrepancy between the two expressions is essentially accounted
for by the Taub-NUT space whose small effects remain at all values of the moduli.
The interesting fact is that the actual micro and macro mechanisms are different.
As we explain below, in the microscopic theory, the Taub-NUT space gives rise to
additional bound states, which changes the degeneracy function, whereas in the
macroscopic formalism, the Taub-NUT space changes the final
value of entropy because of a Chern-Simon coupling in the effective action.  It is
a non-trivial reflection of the consistency of string theory that the two mechanisms in
different regimes of parameter space account quantitatively for the same effect.

\subsection{Microscopic mechanism}

The microscopic setup in type IIB string theory on $K3$ has a D1-D5-p system with the D5 branes wrapping the $K3$, and the effective D1-D5 string with momentum $p$ wrapping a circle $S^1$. The rest of the five dimensions is a KK monopole (Taub-NUT geometry) which asymptotes to $\IR^{3,1} \times \T S^1$. The branes sit at the center of the Taub-NUT space where spacetime looks like $\IR^{4,1}$. The counting of $1/4$ BPS dyons is done by looking at low energy excitations of this system. The counting problem effectively becomes a product of three decoupled systems \cite{David:2006yn} which we can paraphrase as computing the modified elliptic genus of the following 2d SCFT:
\bea\label{effscft}
X^{4d} & = & X^{5d} \times \s(TN_1) \times \s_L(KK-P) \\
X^{5d} & = & \s(Sym^{Q_1Q_5+1}(K3)) \eea

The first factor which is a symmetric product theory which controls
the 5d BPS counting problem of the D1-D5 system. The piece
$\s(TN_1)$ describes the bound states of the center of mass of the
$D1$-$D5$ with the KK monopole. The piece $\s_L(\textrm{KK-P})$
describes the bound states of the KK monopole and momentum and is  a
conformal field theory of $24$ left-moving bosons of the heterotic
string, which can be deduced from the duality between the Type-IIB
KK-P system  and the heterotic F1-P system. The presence of the
second and third factor is crucial for establishing S-duality and
the wall-crossing phenomena in 4d.

The degeneracy of  the BPS states of the theory $X^{4d}$ is given in
terms of the partition function which is the inverse of the Igusa
cusp form, the unique weight $10$ modular form of $Sp(2,\IZ)$.
\begin{equation}\label{invFour}
   \Omega^{4d}(Q_1,Q_5,n,l) = \oint_{\CC}
   d\rho d \s d v \, e^{-2 i\pi \left( \rho n + \s Q_1 Q_5 +  l v \right) }\,
   \frac{1}{\Phi_{10}(\rho, \s, v)} .
\end{equation}
This partition function is understood by separately counting the
three decoupled pieces in the formula (\ref{effscft}) above. The
degeneracies of the theory $X^{5d}$ is given by a similar inverse
fourier transform (\ref{invfive}) with a partition function
$Z(\rho,\s,v)$ which differs sightly from that of the 4d theory.

The discrepancy between the two partition functions
(\ref{ZIgusarel}) is due to the factors $\s(TN_1) \times \s_L(KK-P)$
which completes the 5d system into the 4d system. The BPS partition
function of the extra piece related to the KK monopole is precisely
$f^{KK}(\rho,\s,v)$ (\ref{ZIgusarel}), (\ref{deffKK}). Physically,
most of the entropy of the dyonic black hole comes from the first
factor in (\ref{effscft}) which governs the 5d black hole, but a
small fraction of the entropy of the 4d black hole comes from the
bound states of momentum and center of mass with the KK monopole
itself. This small fraction precisely accounts for the sub-leading
corrections to the 4d-5d lift formula\footnote{In the limit of large charges
in which we evaluate the integral,
the contribution from the KK monopole piece comes purely from the ground state,
and one can explicitly see the equivalence to the macroscopic mechanism
already at this level of the calculation. We thank the referee for pointing this out.}.

\subsection{Macroscopic mechanism}

As we reviewed in section $\S{\ref{macro}}$ the macroscopic entropy
of the black hole is given by Wald's entropy formula. In principle,
one should find the full black hole solution and compute
(\ref{wald}) to obtain the entropy, but in the presence of higher
derivative corrections this can be a very difficult task. For
extremal black holes the attractor mechanism greatly simplifies the
procedure, since only the value of the moduli at the horizon
determines the entropy. Further the entropy function formalism gives
a simplified prescription to evaluate (\ref{wald}) and obtain the
black hole entropy.

In \cite{{Castro:2007hc},{Castro:2007ci}} it was first noted that by
using this procedure the sub-leading corrections to the entropy for
a 4d and 5d black hole differ and the difference is due to a shift
of the charges as given by (\ref{shift}). The shift is a consequence
of the mixed gauge-gravitational Chern-Simons term, where the
curvature of spacetime acts as a source for electric charge.  In the
4d setup, the Taub-NUT space thus effectively absorbs some of the
charge which is placed at the center. If we measure charge using
different Gauss spheres, the measurement near the center of the
taub-NUT space (5d) is different from that at infinity (4d).

\section{Concluding remarks\label{conclusion}}

We would like to finish the discussion by highlighting some of the
implications of our results and future directions  for both the
microscopic and macroscopic approaches.

The microscopic corrections to the black hole entropy that we found
agreed with the macroscopic supergravity theory with higher
derivative corrections.  The
off-shell formalism used to derive such corrections assures that the
action is supersymmetric and insensitive to field redefinitions
because the off-shell algebra does not mix different orders. The
match with the microscopics to this order in $\alpha'$ is therefore
a more stringent test of string theory than in the four dimensional
case.

The equivalence between Wald's formula (\ref{wald}) and the entropy
function (\ref{S5d}) relies on the gauge invariance of the action.
In this case, the conserved charge can be identified from the near
horizon data and it is defined as
\begin{equation}\label{Qf}
Q_I=\frac{\partial f}{\partial e^I}~.
\end{equation}
In order to define the entropy function in the presence of
Chern-Simons terms, one restores gauge invariance by first adding
total derivatives to the action and then dimensionally reducing it
\cite{Sahoo:2006vz}. For black holes on Taub-NUT this procedure will
inevitably
define a four dimensional charge and the effects on the charges
from the delocalized sources due to the curvature of Taub-NUT will
be overlooked. As it stands it seems as if in the presence of the
mixed gauge-gravitational Chern-Simons term a five dimensional
charge cannot be defined using $f$, and there is no extremization
principle. Nevertheless, because of the attractor mechanism,
(\ref{S5d}) evaluated on the solution will determine the same
entropy as defined by Wald's formula. It will be interesting to
determine in the semiclassical theory the appropriate generalization
of the entropy function that will capture delocalized effects and
define an extremization procedure.

On another front, it would be interesting to see if there is a way
to understand -- as for the 4d case -- the entropy of the 5d black
hole in string theory when the various charges in the system are not
equally large (but their product is large). This would necessarily
involve taking into account the corrections due to
worldsheet/membrane instantons which are delocalized in the five
dimensions.

Perhaps the above two questions can be attacked using a generalization of the
entropy function formalism to an integral over paths instead of
minimization of a functional, as suggested in \cite{Sen:2007qy}. It
is possible that the 5d D1-D5-p black hole could be once again be
used as a testing ground for certain fundamental principles in
string theory.

\subsection*{Acknowledgments}

A.~C. would like to thank Joshua Davis, Per Kraus and Finn
Larsen for useful discussions and collaboration on previous work
that motivated the present article. S.~M. would like to thank Nabamita Banerjee,  Edi
Gava, Kumar Narain, Boris Pioline, Bernard de Wit and especially Atish Dabholkar for useful and enjoyable  discussions. S.~M.  would like to thank the hospitality of MCTP, Michigan where this work was initiated and the Monsoon workshop on String theory at TIFR, Mumbai where it was partially completed. The work of A.~C. is supported by
DOE under grant DE-FG02-95ER40899.

\appendix

\section{Appendices}

\subsection{Some details of the evaluation of the contour and saddle point integral \label{integral}}

In this appendix, we shall sketch some relevant details about the
evaluation of the integral (\ref{invfive}) which we recall here.
Consider
\begin{equation}\label{invfiveagain}
   \Omega^{5d}(Q_1,Q_5,n,l) = \oint_{\CC}
   d \T \rho d \T \s d \T v \, e^{-2 i\pi \left( \T \rho n +  \T \s (Q_1 Q_5 +1) +  l \T v \right) }\,
   {Z(\T \rho, \T \s, \T v)} ~.
\end{equation}
The integral above is over the contour
\begin{eqnarray}\label{contouragain}
  &&0 < {\Re(\T \rho)} \leq 1~, \quad 0 < \Re(\T \sigma) \leq 1~, \quad 0 < \Re(\T v ) \leq 1~, \cr
  && {\Im(\T \rho)} \gg1~, \quad \Im(\T \sigma) \gg1~, \quad \Im(\T v )\gg
  1~,
\end{eqnarray}
over the three coordinates, where $\Re$ and $\Im$ denote the real
and imaginary parts. This defines the integration curve $\CC$ as a
3-torus in the Siegel upper half-plane. The imaginary parts are
taken to be large to guarentee convergence. As we shall see below,
the dominant pole in the function is not affected, and we can
therefore perform the contour integral around that pole. This gives
a prescription for the contour. As mentioned in the text, it is
expected that there is no dependence on the moduli in the 5d theory,
and therefore there are no other poles where wall-crossing behavior
occurs in the 5d integral. A precise analysis of the contour as was
done in 4d \cite{Cheng:2007ch} remains to be done.

We mostly follow \cite{Jatkar:2005bh} in the evaluation of the integral. 
First we need to do a contour integral in the $\T v$
coordinate, which picks up the residue at various poles. These poles
occur at zeros of the function $\Phi_{10}$ and the poles of the
function $f^{KK}$. For large charges, the dominant contribution when
the exponent takes its largest value at its saddle point. This was
analyzed in \cite{Dijkgraaf:1996it}. When $f^{KK}$ is not present, this dominant divisor
is
\be\label{divisor} \T \rho \T \s - \T v^2 + \T v =0~. \ee
We can check that the function (\ref{deffKK})
\be\label{fkkagain} f^{KK}(\rho,\s,v)  = p \, \eta^{18}(\rho) \,
\vartheta_1^2(v, \rho)~,  \ee
does not take away this pole, and does not alter the dominance of
this pole. We can now carry out the contour integration in the
variable $\T v$ around the zero of the above divisor
\be\label{vsol} \T v_{\pm} = \frac{1}{2} \pm  \Lambda( \T \rho, \T \s)  \, , \qquad   \Lambda( \T \rho, \T \s)  = \sqrt{\frac{1}{4} + \T
\rho \T \s}~. \ee
In the contour integration, the variables $\T \rho$ and $\T \s$ are
held fixed and we choose the negative value of the square root $\T
v_{-}$.

The modular properties of the function $\Phi_{10}$ under $Sp(2,\IZ)$
allow us to factorize it around the value $\T v = \T v_{-}$. The
integrand in (\ref{invfive}) behaves like:
\be\label{factorize} C \exp \left(- 2 \pi i (\T \rho n + \T \s (Q_1
Q_5 +1) + 2 \T v l ) \right) \T \s^{12} (\T v - \T v_{+})^{-2} (\T v
- \T v_{-})^{-2}  \eta^{-24} (\rho)   \eta^{-24} (\s) f^{KK}(\T
\rho, \T \s, \T v)~. \ee
Using this factorization, we can evaluate the contour integral, and
then perform a saddle point analysis of the remaining integral over
$(\T \rho, \T \s)$.  The contour integral for $\T v$ gives
\be\label{inteval}
 \Omega^{5d}(Q_1,Q_5,n,l) = (-1)^{Q.P}  K \int d \T \rho \, d \T  \s \, e^{X(\t \rho, \T \s) + \ln \Delta(\t \rho, \T \s)~,}
\ee
where $K$ is a numerical constant and 
\bea\label{defX}
X (\t \rho, \T \s) & =   &- 2 \pi i \left( \T \rho n + \T \s (Q_{1} Q_{5}+1) -  \Lambda (\T \rho, \T \s) l\right) \cr&& +12 \ln{\T \s}  
- \ln \eta^{24}(\rho) - \ln \eta^{24}(\s) + \ln f^{KK}(\T \rho, \T \s, \T v_{-}) ~, 
\eea
\bea\label{defDelta}
\Delta(\t \rho, \T \s) & = & \frac{1}{4 \Lambda( \T \rho, \T \s)^{2}} \left[ - 2 \pi i l+ 2 \frac{\T v_{-}}{\T \s} \frac{\p}{\p \rho} \ln \eta^{24}(\rho) - 2 \frac{\T v_{+}}{\T \s} \frac{\p}{\p \s}\ln \eta^{24}(\s)  \right. \cr
&& \qquad \qquad  \qquad \left.  + \frac{1}{\Lambda( \T \rho, \T \s)} +  \left( \frac{\p}{\p \T v} \ln f^{KK}(\T \rho, \T \s, \T v) \right)_{\T v = \T v_{-}}  \right] \ .
\eea
The above expression has to be evaluated at the saddle point. In the large charge limit
\begin{equation}\label{Qlarge}
Q_1Q_5\gg 0~, \quad n\gg 0~, \quad \sqrt{Q^3-J^2}\gg0~,
\end{equation}
the saddle point of \eqref{inteval} is well approximated by the first line of \eqref{defX}, and in this limit it is located at
\begin{equation}\label{app:saddle}
\T \rho  =  \frac{i}{2} \frac{Q_1 Q_5
+1}{\sqrt{Q^3 - J^2}}~, \quad \T \s  =  \frac{i}{2}
\frac{n}{\sqrt{Q^3 - J^2}}~,
\end{equation}
which is the extremum given by \eqref{rsvvalues}. 
We can now estimate the above expressions \eqref{defX}, \eqref{defDelta} for the two relevant limits used in $\S\ref{sec:saddle}$: the Supergravity regime, {\it i.e.} $Q_{1} Q_{5} \sim N^{2}$, $n \sim N$ and $l \sim N$; and the Cardy regime, {\it i.e.}   $Q_{1} Q_{5} \sim N$, $n \sim N^2$ and $l \sim N$ with $N\gg 1$. For both regimes, \eqref{defX} evaluated at \eqref{app:saddle} behaves as 
\begin{equation}
X(\T \rho, \T \sigma)= {\rm (const)} N^{3/2}+{\rm (const)} N^{1/2}+  {\cal O}(1)~,
\end{equation}
where the precise values of the constants are computed in section $\S\ref{sec:saddle}$ for each regime. Next, the subleading behavior of   \eqref{defX}, \eqref{defDelta} relevant for the saddle point approximation are
\bea\label{Deltaest}
\ln \Delta & = & - \ln |l| + \ln(\frac{1}{4} + \T \rho \T \s) + {\cal O}(1) \ , \cr
\ln \left({\rm det}|\partial^2 X|\right)  & = &  \ln |l| - \ln(\frac{1}{4} + \T \rho \T \s) + {\cal O}(1) \ ,
\eea
where $\partial^2X$ is the matrix of second derivatives of $X$ with respect to $\T\rho$ and $\T\sigma$. Here, ${\cal O}(1)$ refers to the above large charge expansion, and refers to the scaling as a function of $N$. Finally, integrating \eqref{inteval} using the saddle point approximation, the statistical entropy is given by
\begin{eqnarray}
S^{5d}_{stat} &=&\ln\left( \Omega^{5d}(Q_1,Q_5,n,l) \right) \nonumber\\
& = & - 2 \pi i {\T \rho} n - 2 \pi i \T \s
(Q_1 Q_5 +1) + 2 \pi i (\frac{1}{2} - \t v) l \nonumber\\ & 
&+ 12 \ln \T \s - \ln \eta^{24}(\rho) - \ln \eta^{24} (\s) + \ln
f^{KK}(\T \rho, \T \s, \T v)+{\cal O}(1)~, 
\end{eqnarray}
evaluated at \eqref{app:saddle}. From the above analysis, $S_{stat}^{5d}$ allows  a systematic expansion  for both the Supergravity and Cardy regime. 


Note that the function $f^{KK}$ does not have any poles in the
interior of the region we are considering, but has many zeroes.
These zeroes do not include the divisor (\ref{divisor}).
Therefore the dominant pole of $\Phi_{10}^{-1}$ remains the dominant
pole of the 5d integrand $Z$. Note however that $f^{KK}$ does have a
zero at $\T v = 0$ which takes away the pole at the same value of
the function $\Phi_{10}^{-1}$. This means that there is no wall
crossing behaviour in the five dimensional theory due to this pole.
For the evaluation of the integral, these observations mean that the
presence of the function $f^{KK}$ changes the analysis only through its
appearance in the entropy function (\ref{sstat}) to be extremized.

\subsection{The Jacobi $\eta$ and $\vartheta$ functions and their properties \label{jacobi}}

We define
\be\label{defqy} q = e^{2 \pi i \tau}~, \quad y = e^{2 \pi i v}~.
\ee
The Jacobi eta function is defined as
\be\label{defeta} \eta(\tau) = q^{\frac{1}{24}} \prod_{n=1}^{\infty}
(1 - q^n) ~. \ee
The odd Jacobi theta function is
\be\label{deftheta} \vartheta_1(v,\tau) = - 2 q^{\frac{1}{8}} \,
\sin(\pi v) \prod_{m=1}^\infty (1-q^m) (1-q^m y) (1-q^m y^{-1}) ~.
\ee
For large imaginary values of $\tau = it,\, t \to \infty$, we have
$q \to 0$ most of the terms in the product become unity and these
functions admit an expansion of the form
\be\label{etaexp} \eta(\tau) = - \frac{\pi}{12} t + \ldots \ee
These functions satisfy the modular properties:
\bea\label{modulareta} \eta(-\frac{1}{\tau}) & = & \sqrt{-i \tau}
\eta(\tau) \cr \vartheta_1(\frac{v}{\tau},-\frac{1}{\tau}) & = & i
\sqrt{-i \tau} e^{i \pi v^2/\tau} \vartheta_1(v,\tau)~. \eea
For the $\vartheta$ function, the expansion depends on the value of $v$ compared to $\tau$, but similar expansions are possible.

\subsection{The Jacobi-Rademacher expansion \label{Radem}}

The Jacobi-Rademacher expansion
\cite{{Dijkgraaf:2000fq},{Manschot:2007ha}}  is a very powerful
(exact) expansion containing  both power law and exponential
corrections to the Cardy estimate. Here, we are only interested in
the first power law correction, which can be estimated by using a
Jacobi modular transformation and a saddle point expansion.

The counting of $1/4$ BPS states of the D1-D5 system on $K3$ is
summarized by the elliptic genus of the 2d SCFT $Sym^k(K3)$ with $k
= Q_1Q_5 +1$. This elliptic genus can be expanded in a theta
function decomposition
\bea\label{thetadecom}
\chi(Sym^k(K3);\tau, z) & = & - \sum_{l = -k+1}^{k} \sum_{n \in \IZ} c(n,\mu) \, q^{n - l^2/4k} \, \theta_{l,k}(z,\tau)\\
& \equiv  & - \sum_{l = -k+1}^{k} h_l (\tau) \,
\theta_{l,k}(z,\tau)~. \eea
We write
\be\label{Hexp} h_l(\tau) = \sum_{m=0}^\infty H_l(m) \, q^{m -
\frac{l^2}{4k}} \, . \ee
We can estimate the value of the coefficients $H_l(n)$ when $n \gg
k$ using the Cardy's formula after doing a modular transformation on
the elliptic genus and performing a saddle point expansion
\be\label{Cardysaddle} H_l(n) = (const) \, e^{\pi i l} \frac{k}{(4 n
k - l^2)^\half} \, I_{3/2}(2 \pi \sqrt{n k - l^2/4}) + \ldots~, \ee
where the dots denote terms which are exponentially suppressed.
There is actually an exact formula which captures all the
exponentially sub-leading terms
\cite{{Dijkgraaf:2000fq},{Manschot:2007ha}} which we don't need
here.

Here $I_{3/2}$ is the modified Bessel function of the first type.
The index $3/2$ appears because the weight of the vector valued
modular form $H_\mu(z)$ is $w=-\half$. Note that by definition, the
elliptic genus has weight zero, but the $\theta$ functions have
weight $+\half$, so the functions $H_\mu$ have weight $-\half$. This
function in fact has an expression in terms of elementary functions
\be\label{Bessel3half} I_{3/2} (z) = \sqrt{\frac{2}{\pi z}}
\left(\cosh(z) - \frac{\sinh(z)}{z} \right)~. \ee
The entropy is the logarithm of the degeneracy $H_\mu(n)$. With $k =
Q_1 Q_5 +1$, we have $z=2 \pi\sqrt{(Q_1Q_5 +1)n-{\ell^2}/4}$. The
entropy is equal to
\bea\label{entropyexp}
S^{5d} & = & \ln \left(e^z\left[1 - \frac{1}{z}\right] \right) + \ldots \\
& = & 2 \pi \sqrt{(Q_1Q_5 +1)n-{l^2}/4} \left(1  + \frac{1}{4\pi^2
(Q_1Q_5n-{l^2}/4)}  + \ldots\right)~, \eea
which is in agreement with (\ref{entropylim3}).

\bibliographystyle{utphys}
\bibliography{all}

\end{document}